\documentclass[fleqn,pdflatex,english]{article-hermes}

\proceedings{VIRA}{-}
\title[High resolution imaging with VISIR]%
      {High resolution mid-Infrared imaging of dust disks structures around Herbig Ae stars with VISIR}

\author{Coralie Doucet\fup{*} \andauthor Pierre-Olivier Lagage\fup{*} \andauthor Eric Pantin\fup{*}}
\address{%
\fup{*} AIM, Unit\'e Mixte de Recherche CEA - CNRS - Universit\'e Paris VII -UMR 7158\\  DSM/DAPNIA/Service d'Astrophysique\\ CEA/Saclay, F-91191 Gif-sur-Yvette, France\\[3pt] doucetc@cea.fr\\}

\resume{Nous présentons un nouveau mode d'observation avec VISIR, l'instrument pour les observations en InfraRouge (IR) moyen au VLT (ESO, Chili): le \textbf{mode BURST}. Ce mode permet d'atteindre la limite de résolution spatiale déterminée par la diffraction. Une illustration d'utilisation de ce mode est donnée avec l'observation des disques autour d'étoile de type Herbig Ae. 
Les fen\^etres atmosphériques en IR moyen sont appropriées à l'étude des extensions autour de ces objets. En effet, avec un télescope de 8 mètres, il est possible d'atteindre la limite de diffraction de 0.3 seconde d'arc dans de bonnes conditions de seeing. Ainsi, on peut résoudre des disques d'une taille typique de 100 AU pour des étoiles situées à 100 pc. HD97048 est un remarquable exemple d'une étoile Herbig Ae avec un disque évasé imagé à 11.3 $\mu$m (PAH). }

\abstract{We present a new mode of observations with VISIR, the mid-InfraRed (mid-IR) imager and spectrometer on the VLT (ESO, Chile): the so-called \textbf{BURST mode}. This mode allows to reach the diffraction limit of the telescope. To illustrate results obtained with this mode, we discuss observations of disks around Herbig Ae stars.
The 10-20 $\mu$m atmospheric windows are well-suited to study the extended emission of these objects. With a 8 m class telescope, in fair seeing conditions, the observations are diffraction-limited at 10 $\mu$m and the spatial resolution could reach the diffraction limit of 0.3 arcsec. As a result, it is possible to resolve disks with a typical size of 100 AU around objects at a distance of 100 pc. We present here a significant example, HD97048, for which a flared disk of 350 AU is resolved at 11.3 $\mu$m (PAH band).}

\motscles{étoiles Herbig Ae, HD97048, BURST mode, VISIR, résolution angulaire}


\begin{document}


\maketitlepage

\section{Introduction}

Circumstellar (CS) disks made of gas and dust are ubiquitous around young stars.  They are a natural
  outcome of the star formation process, because of the need of
  angular momentum conservation during the collapse of the initial
  molecular core \cite{SHU87}.
   As the star evolves, the disk changes: the gas is dissipated and a variety of processes (such as collisions for instance) leads to the growth of dust grains and eventually to the formation of planets. In order to study planets formation, 
   it is necessary to understand the physics of the medium where they were born.
   Herbig Ae (HAe) stars represent a particularly interesting
  laboratory for studying disks evolution and planet formation. They are believed to be the more massive analogues of T Tauri
  stars ($\sim$ 2-10 M$_{\odot}$), and harbour disks in which planets are still in the building process or eventually just formed. Although great progresses have been made in modelling the disk structure with radiative transfer codes able to reproduce the 
  Spectral Energy Distribution (SED) [\cite{CG97}, \cite{Natta01}, \cite{DDN01}], the structure of the disks is not uniquely constrained.
  Spatially resolved images of the disk are needed to better constrain the disks. 
The SED of about half of the HAe stars features IR emission bands \cite{ACKE04b} which are believed to be produced by very large organic molecules (like PAH: polycyclic aromatic hydrocarbon particles). Such particles are transiently heated by star light UV photons: they do not reach thermal equilibrium with the radiation 
field, but absorb individual photons, experiencing a rapid increase in temperature, and then cool down, 
re-radiating the absorbed energy in IR emission bands. Such a radiation allows to probe, in the mid-IR, the disk structure at large distances from the star. Furthermore, observations in the mid-IR, with a 8 m class telescope, in fair seeing conditions, can be diffraction-limited, so that the spatial resolution can reach 0.3 arcsec. As a consequence, it is possible to resolve disks with a typical size of 100 AU around Herbig stars located at a distance of 100 pc.

\section{New imaging mode of VISIR: the BURST mode}
The observations were performed using the ESO mid-infrared instrument VISIR installed on the VLT (Paranal, Chile). The instrument is equipped with a DRS (former Boeing) 256x256 pixels BIB detector array.


   
\begin{figure*}[!h]
\begin{minipage}[t]{5.8cm}
\centerline{\resizebox{5.7cm}{!}{\includegraphics[angle=0]{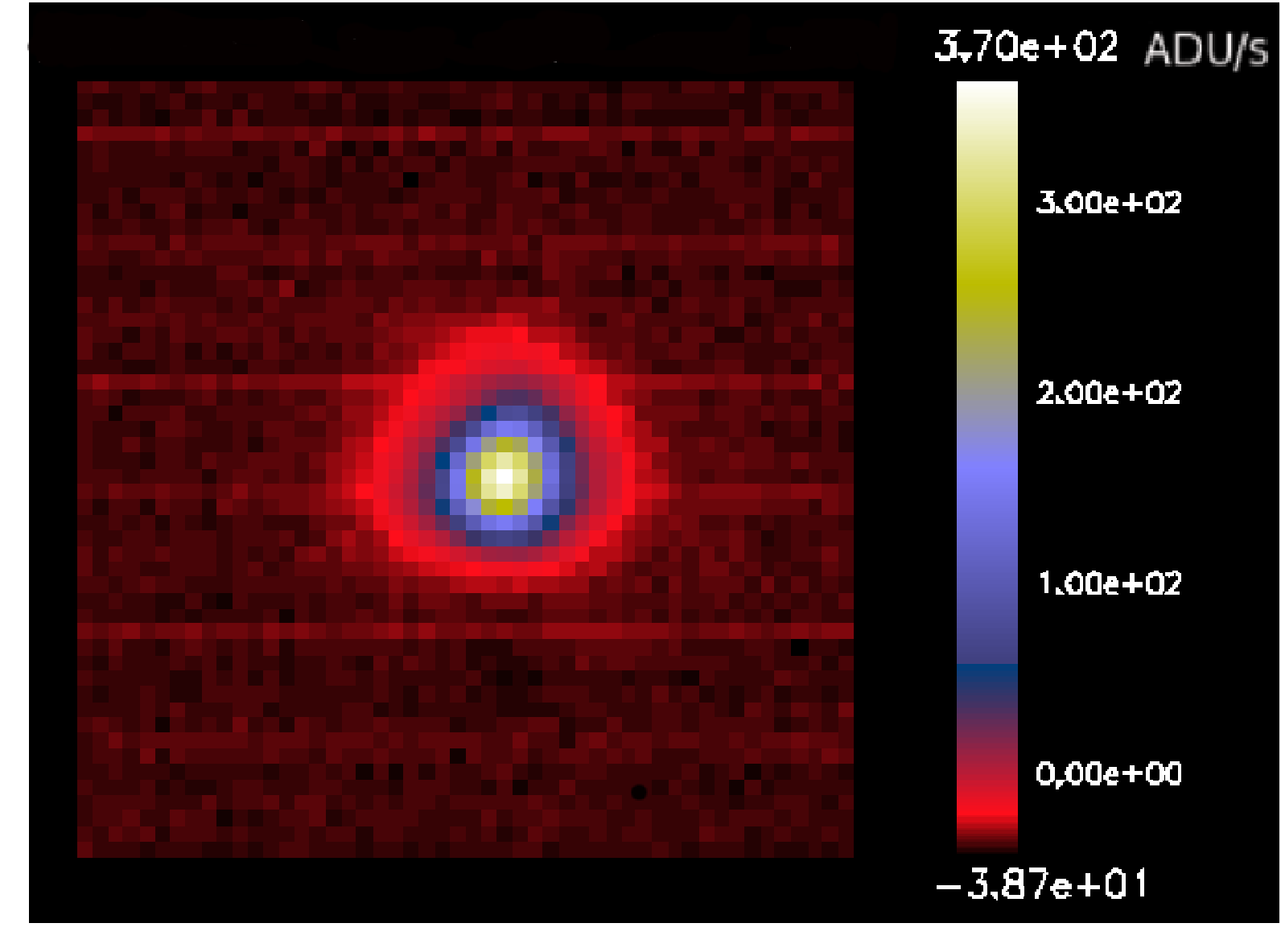}}}
\end{minipage}
\begin{minipage}[t]{5.8cm}
\centerline{\resizebox{5.7cm}{!}{\includegraphics[angle=0]{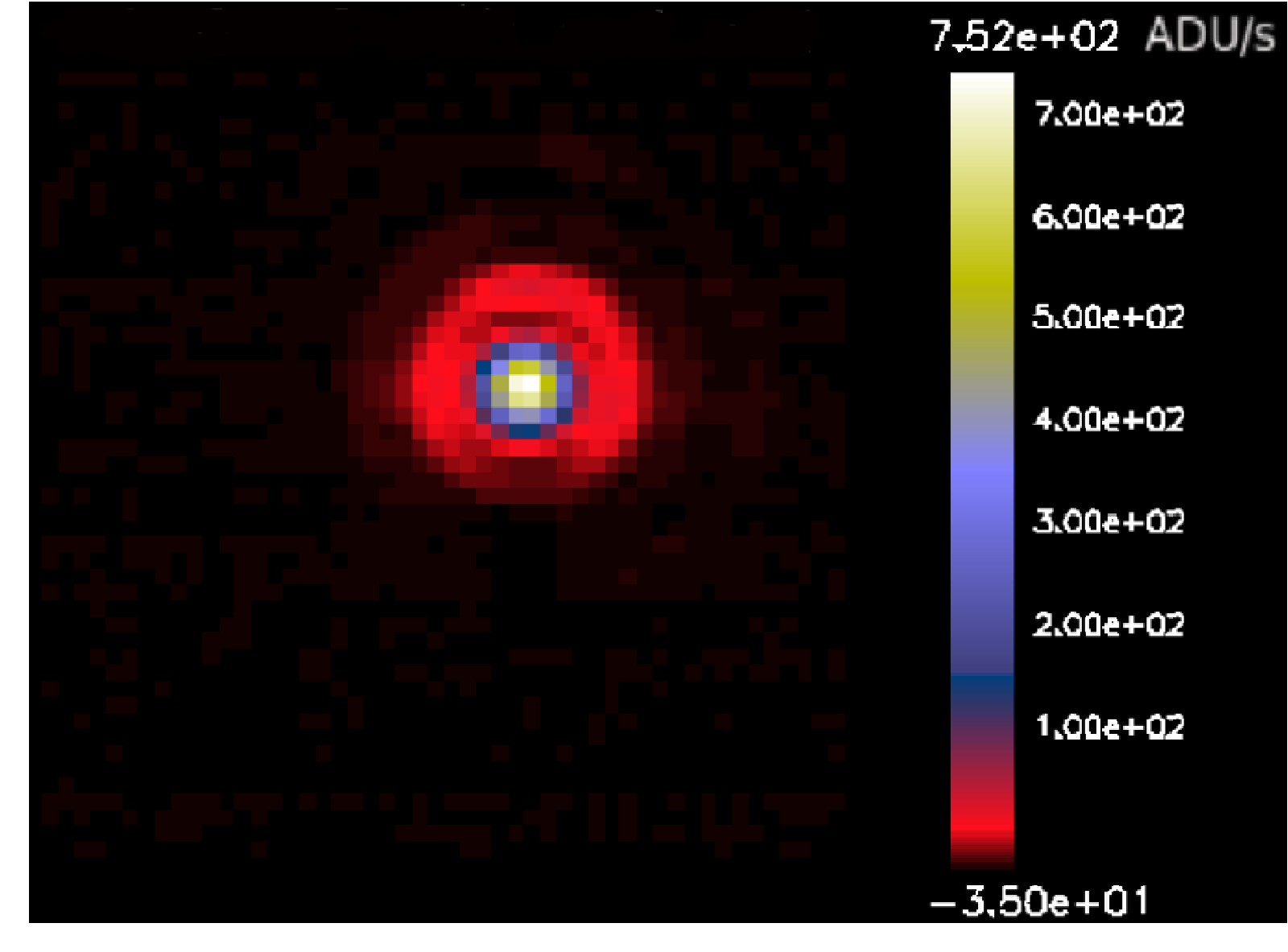}}}
\end{minipage}
\caption{VISIR images of a reference star resulting from adding 429 chopped/nodded images with an elementary exposure time of 50 ms. The left image is a direct sum of the images in the cube; the right image is built using the autocorrelation shift and add procedure. The Signal over Noise ratio has been increased by more than a factor 2 by this proedure (sensitivity of 17 mJy/10$\sigma$/1h in the "raw" left image and of 7 mJy/10$\sigma$/1h in the shift-and-added  image). }\label{comparaison_burst}
\end{figure*}

\begin{figure}[!h]
\begin{center}
\includegraphics[angle=90,scale=0.32]{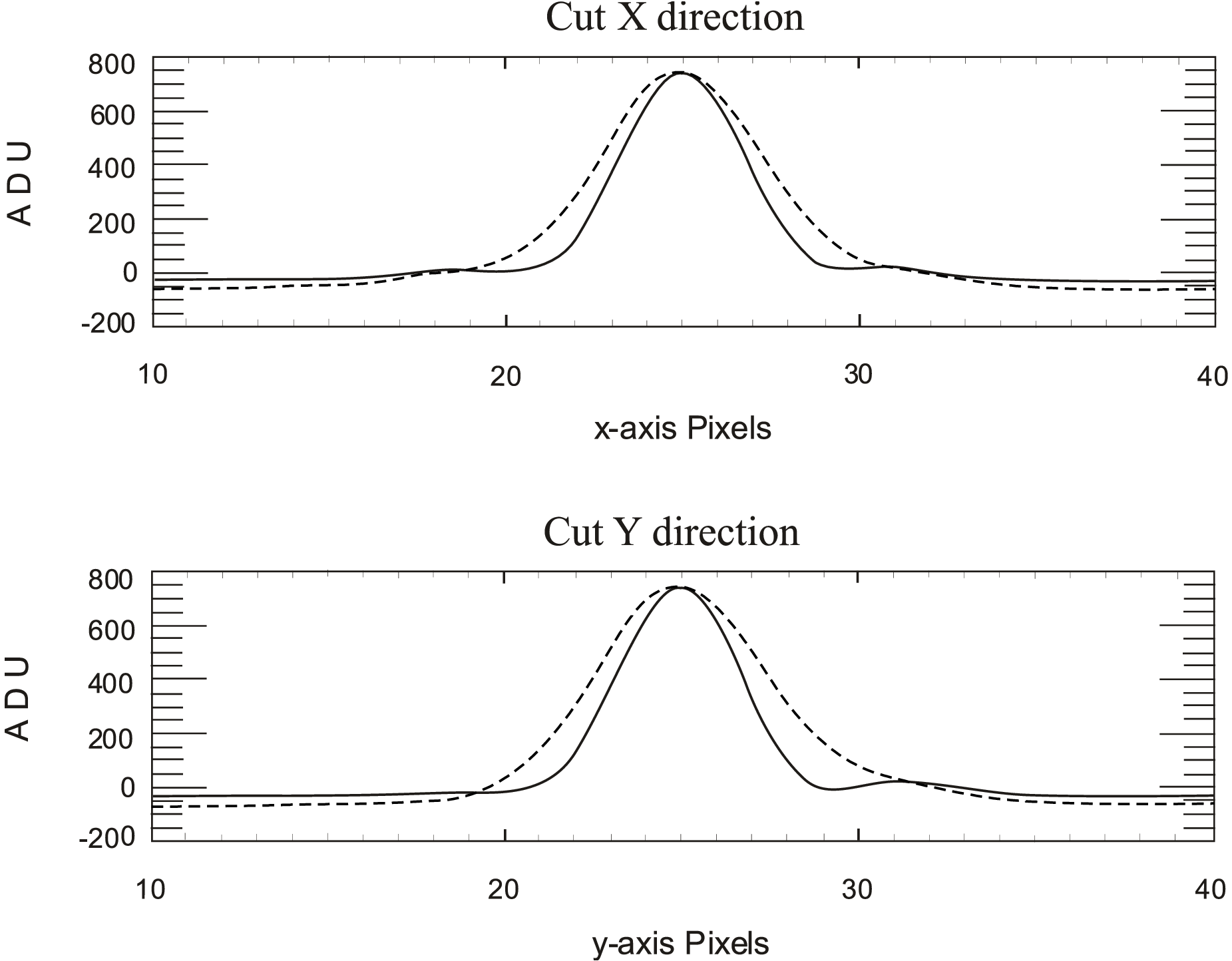}
\caption{Comparison of the profile on the two images of Fig~\ref{comparaison_burst}. The non shift-and-add corrected image (dashed line) has a FWHM 1.5 times larger than the corrected image (plain line).}\label{comparaison_qualite_image}
\end{center}
\end{figure}

Under good seeing conditions($\leq$0.5 arcsec in the visible), the images in the mid-IR are diffraction-limited even on a 8 meters class telescope. Unfortunately, the median seeing experienced at Paranal is of the order of 0.8 arcsec, which degrades significantly the angular resolution. Indeed, for a seeing of 0.8 arcsec in the visible, the seeing value at 10 $\mu$m is 0.4 arcsec, when assuming that the wavelength dependence of the seeing follows a $\lambda^{-1/5}$ law. This is larger than the diffraction limit of 0.3 arcsec and represents a 5 pixels movement on the detector with the smallest field of view of VISIR (0.075"/pixel). In order to get the best
spatial resolution with VISIR, we experimented a new imaging mode on bright objects: the \emph{BURST mode}. The
principle is to take short enough exposure images ($\lesssim$ 50 ms) in order to freeze the turbulence; the coherence time of the atmosphere at 10 $\mu$m is around 300 ms at Paranal for a good seeing. But as soon as the exposure time is short, speckles appear with the diffraction spot. 
 The number of speckles is linked to the ratio $D/R_{0}$ where $R_{0}$ is the Fried parameter or coherence length and D the telescope aperture. Since the Fried parameter is the order of the diameter of the telescope in the mid-IR (Table~\ref{turbulence}), i.e much larger than in the near-IR, the number of speckles in the mid-IR is small and one can have one principal speckle in the image for good seeing conditions. 
 In order to correct for the turbulence by offline processing, we stored 1000 elementary images by nodding position for a chopping frequency of 0.25 Hz in the direction north/south. The nodding direction is perpendicular to the chopping direction with an amplitude of 8". After classical data reduction in mid-IR, a cube of 500\footnote{1000 divided by 2 because of the 2 chopper positions} images chopped and nodded (4 beams/image) is obtained. Because of the turbulence, each source on an image moves independently and as a result, we have to extract individually the 4 sources in each image (4 quarters) of the cube and shift and add the image with the ones corresponding to the same quarter. Finally, we shift and add the four final images of the four quarters (Fig.~\ref{comparaison_burst} and Fig.~\ref{comparaison_qualite_image}).

\begin{table}[!h]
 \begin{center}
 \begin{tabular}{c|c}
 \hline
               wavelength ($\mu m$)      &  R$_{0} (m)$    \\               
\hline
  \hline
  0.5    &  $\sim$ 0.15  \\
 8.6  (in this study)&   6.23 $\pm$ 0.33\\
 11.3 (in this study)& 8.76 $\pm$ 0.42 \\        
\hline
 \end{tabular}
\caption{Comparison of the Fried parameter $R_{0}$ of the atmosphere for different wavelengths. The Fried parameter for mid-IR has been calculated in this study with measuring the angle-of-arrival fluctuation for a point source observed in the image plane of the telescope (Brandt et al., 1987).}
 \end{center}\label{turbulence}
\end{table}

\section{Image quality and shift and add method}
\begin{figure*}[!h]
 \begin{center}
 \includegraphics[angle=0,scale=0.27]{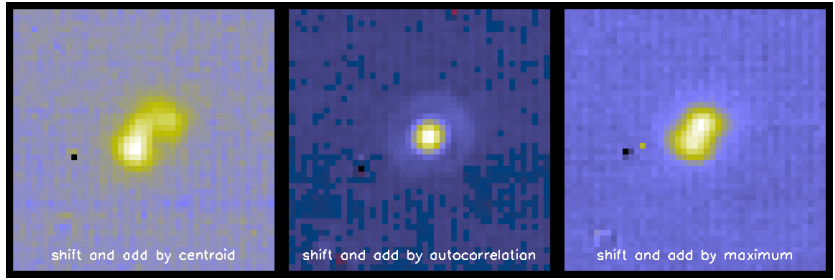}
 \caption{For a visible seeing of 1.2 arcsec, comparison of the image quality when shift and add is performed using maximum finding, centroid estimate or
 autocorrelation method.}\label{bad_seeing}
 \end{center}
 \end{figure*}

As it has already been done in the near-IR \cite{CHRI91}, we investigate the image quality in mid-IR by experimenting three differents methods of shift and add in which the offsets are calculated using a centroid estimate, a maximum finding or an autocorrelation method. For a good seeing, typically 0.75 arcsec, the method of maximum and centroid are comparable and give poor quality results (see Table~\ref{comparaison_method_shift_tab}); the autocorrelation method allows to recover diffraction-limited images in N band.
As the seeing degrades (Fig.~\ref{bad_seeing}), the autocorrelation method consistently yields improved resolution maintaining diffration-limited images, at the expense of rejecting more images (40 \% of the images for a seeing of 1 arcsec).

\begin{table}[!h]
\begin{center}
 \begin{tabular}{l|c|c|c}
 \hline
   & centroid  & autocorrélation & maximum  \\
   & estimate & method                 & detection \\
  \hline
  \hline
  FWHM$_{x}$ (mas)  & 424 $\pm$ 20  & 302 $\pm$20 & 386  $\pm$20    \\
  FWHM$_{y}$ (mas)  &  547 $\pm$20   & 288  $\pm$20  & 523 $\pm$20   \\
  \hline
  ellipticity &      1.29   &       0.95 &        1.36\\
  \hline
  Strehl ratio& 0.329 $\pm$ 0.002 & 0.368 $\pm$ 0.001 & 0.760 $\pm$ 0.001 \\
 \hline
 \end{tabular}
 \caption{Comparison of the FWHM in milli-arcsec (mas) with the theorical one at 11.3 $\mu$m of 283 mas. Strehl ratio for the final image obtained with the shift and add executed by centroid estimate, maximum finding or autocorrelation method (for a visible seeing of 0.75 arcsec).}\label{comparaison_method_shift_tab}
 \end{center}
 \end{table}

\section{Limits of the method}
This mode could only be used for objects bright enough so that the signal over noise in an elementary frame is high enough to apply the shift and add procedure (Flux > 5 Jy in N band and 10 Jy in Q band) with reasonable seeing conditions (less than 1.3 arcsec in the visible). Furthermore, it produces a huge amount of data since one  hour of observations produces 7 Gb, taking into account large overhead, and if the efficiency would be increased, then the rate could increase up to 30 Gb per hour.

 \section{HD97048, an outstanding example}
 \begin{figure*}[!h]
 \begin{center}
 \includegraphics[angle=0,scale=0.35]{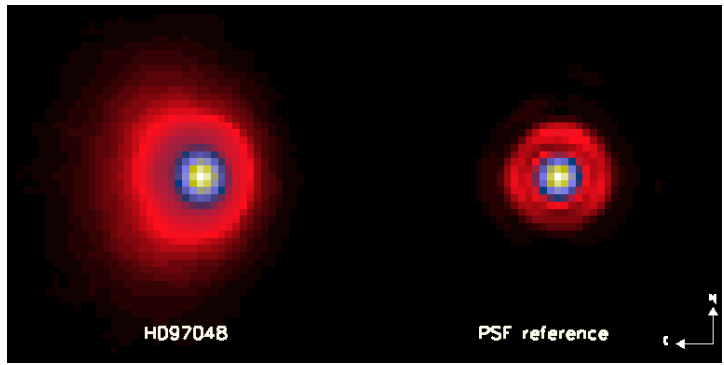}
 \caption{HD97048 (on the left) taken in BURST mode at 11.3 $\mu$m (PAH band). The extension is quite large (up to 300 AU) as compared to the reference star (on the right) and asymmetric in the east/west direction.}\label{ima}
 \end{center}
 \end{figure*}
 
 HD97048 is a Herbig Ae star located at 180 pc in the Chameleon cloud. Figure~\ref{ima} shows that HD97048 is quite extended in PAH (11.3 $\mu$m) compared to the reference star with the same filter. The most striking result is the asymmetry in the wings: the east side is much more extended (around 1.65 arcsec) than the west side (around 1 arcsec).
 This asymmetry (the fact that the bright source is not in the middle of the extension) reveals a disk optically thick and geometrically thick. PAHs allow to detect disks in the mid-IR range further away from the star since they are stochastically heated by UV radiation in a flared disk. They are new tracers of the geometry of the disk surface.
 

\section{Conclusions and perspectives}
The \textbf{BURST mode} has shown its capability to improve image quality when observing from the ground in the mid-IR. Thanks to the delivery of diffraction-limited images, it allows us to spatially resolve disks around Herbig Ae stars. Many other programs requiring the best achievable spatial resolution would benefit from such a mode. We recommend this mode to be implemented as a VISIR standard oberving mode and to be offered to the observers.



\bibliography{bibliographie_astroph}



\end{document}